\documentclass[11pt]{article}
\usepackage{putex}
\usepackage[colorinlistoftodos]{todonotes}
\usepackage{tcolorbox}
\usepackage{amsfonts}
\usepackage{amssymb}
\usepackage{amsmath}
\usepackage[bookmarks = false, colorlinks = true, linkcolor = blue, citecolor = purple]{hyperref}
\usepackage{ulem}
\usepackage{tipa}
\usepackage{graphics,graphicx}
\usepackage{setspace}
\numberwithin{equation}{section}

\newcommand{\HH}{\mathcal{H}}

\newcommand{\VV}{\mathcal{V}}

\newcommand{\AAA}{\mathcal{A}}

\newcommand{\BB}{\mathcal{B}}

\newcommand{\Tr}{\mbox{Tr}}

\newcommand{\OO}{\mathcal{O}}

\newcommand\be{\begin{equation}}
\newcommand\ba{\begin{eqnarray}}
\newcommand\ee{\end{equation}}
\newcommand\ea{\end{eqnarray}}

\definecolor{lred}{rgb}{0.8,0,0}
\definecolor{lurple}{rgb}{0.8,0.0,0.6}
\definecolor{purple}{rgb}{0.5,0.0,0.3}
\definecolor{darkblue}{rgb}{0.0,0.0,0.7}
\definecolor{darkgreen}{rgb}{0.0,0.35,0.0}
\definecolor{huh}{rgb}{0.0,0.6,0.8}
\definecolor{lorange}{rgb}{1,0.5,0}
\definecolor{orange}{rgb}{0.8,0.4,0}
\definecolor{dorange}{rgb}{0.6,0.2,0}
\definecolor{light-gray}{gray}{0.5}

\begin{document}
\title{
Ryu-Takayanagi area from Virasoro modular data 
}
\authors{Jennifer Lin\footnote{e-mail: \texttt{jylin04@gmail.com}}}
\institution{PIBBSS}{Principles of Intelligence, USA }
\abstract{\noindent 
We show that in holographic 2d CFTs, the entanglement entropies across several choices of global state and subregion 
can be written in a way that at once has a microscopic interpretation 
and matches the leading large$-c$ organization of the Ryu-Takayanagi formula. 
This representation is obtained by applying crossing symmetry to the replica manifolds.
From the boundary point of view, each rewritten entropy looks like an algebraic entanglement entropy for the Virasoro algebra restricted to the region, with center labels obtained by coarse-graining heavy primaries of the BCFT on the regulated region into bins labeled by Liouville momenta.
At large $c$, a resulting sum over bins is dominated by a saddle, and the $O(c)$ part of the entropy comes from the Cardy density of heavy primaries in the dominant bin. We identify this $O(c)$ part of the entropy with the Ryu-Takayanagi area. 
Physically, this suggests a concrete statistical origin for the Ryu-Takayanagi area as coming from coarse-grained Virasoro intertwiners across the entangling cut.  
The result also provides quantitative criteria for the amount of coarse-graining allowed for the consistency of the interpretation. 
} 

\maketitle
\newpage

\tableofcontents 

\section{Introduction}
In recent years, it has been suggested that there may be a deep connection between gravity and  entanglement. This idea is most concretely manifested in the Ryu-Takayanagi (RT) formula \cite{Ryu:2006bv, Ryu:2006ef, Faulkner:2013ana} 
	which states that for a region $A$ in a state of a holographic CFT with a geometric bulk dual, the entanglement entropy of $A$ can be expressed in terms of the minimal-area bulk surface $\gamma_A$ homologous to it as
\be\label{rt}
S(A) = \frac{\mbox{Area}(\gamma_A)}{4G_N} + S_{bulk} + \dots\,.
\ee
Here $S_{bulk}$ is the entanglement entropy of bulk quantum fields across $\gamma_A$. 
Although \eqref{rt} has passed many checks and been proved using Euclidean gravity techniques \cite{Lewkowycz:2013nqa}, we don't yet understand why it is true from a canonical point of view. We would like to understand which boundary degrees of freedom account for the area term in particular and how they organize to give 
	the bulk geometry. 
	
In this work, we present an observation that may help with this problem for the special case of AdS$_3$/CFT$_2$. Namely, we show that across several choices of global state and subregion in a holographic 2d CFT (including two intervals in the vacuum state and a single interval in a class of excited state produced by the insertion of a local operator), one can write the entanglement entropy to leading order in $c$ in a form that at once suggests which microscopic degrees of freedom may contribute to the entropy and isolates an $O(c)$ contribution that we identify with the Ryu-Takayanagi area. \footnote{Up to an assumption about the large$-c$ scaling of a certain ratio of conformal blocks that we flag throughout the text.} We can therefore use this representation of the entropy to translate between the RT area and a would-be microscopic explanation for it.

On the boundary side, 
each rewritten entropy 
will have the form of an algebraic entanglement entropy for an algebra with a  nontrivial center,
 containing a Shannon term for would-be center labels, an intra-sector entropy and a sector multiplicity.
 By examining the individual terms of this decomposition, we will argue that a physical candidate for a would-be underlying algebra is the Virasoro subalgebra in region $A$, with center labels obtained by coarse-graining heavy primaries in the BCFT supported on $A$ into bins labeled by Liouville momenta. 
 At large $c$, a resulting sum over superselection sectors is dominated by a saddle, and the 
 $O(c)$ part of the entropy comes from the Cardy density-of-states part of the dominant sector multiplicity. Hence, the RT area arises microscopically as this density-of-states part of the dominant sector multiplicity from this point of view.

More precisely:

\begin{enumerate}	
\item {\bf Crossed-channel formula.} We first show that for a region $A$ in a state $|\psi\rangle$ of a large$-c$ 2d CFT where the criteria for vacuum block dominance \cite{Hartman:2013mia} are valid, one can rewrite the entanglement entropy to leading order in $c$ as 

\be \label{rc}
S(A,|\psi\rangle) = \min_{\rm channels} \int d\vec{P}\, p_{\vec P} (-\log p_{\vec P} + \sum_i \log S_{P_i1} + S_{IR}(\vec P))\,,
\ee
where
\begin{itemize}
\item[*] the $S_{P_i1}$'s, with $i$ running over the number of disconnected components of $A$, are modular S-matrix elements 	
\be\label{sp1}
	S_{P1} = 4 \sinh (2\pi bP) \sinh (2\pi b^{-1} P)
\ee
for $b$ defined via $c = 1+6Q^2$ and $Q = b+b^{-1}$;
\item[*] 
\be\label{sbulk}
	S_{IR}(\vec P) = -\frac{\partial_n\mathcal{F}_{\vec P}(n)|_{n \rightarrow1}}{\mathcal{F}_{\vec P}(1)} + \log \mathcal{F}_{\vec P}(1)
\ee
 has the schematic form of a replica entropy itself, for $\mathcal{F}_{\vec P}(n)$ a particular conformal block on the $n$-replicated manifold, with internal representations $\vec{P}$;
 \item[*] $p_{\vec P}$ is a quasi-probability distribution, schematically a product of the $S_{P_i1}$'s and the block $\mathcal{F}_{\vec P}(1)$ (which in turn depends on both $A$ and $|\psi\rangle$), that we will define case-by-case below;
\end{itemize}
and we minimize over a choice of channel for the $\mathcal{F}_{\vec P}(n)$'s.

Eq. \eqref{rc} is obtained by applying crossing symmetry on the replica manifold.
Previously, a formula like it was written down for the entanglement entropy of a single interval in the vacuum state of a 2d CFT \cite{Wong:2022eiu}, \cite{Lin:2021veu}.
In this work, we generalize the earlier result in two directions: to two intervals in the vacuum state of a large$-c$ 2d CFT, and to a single interval in an excited state of a large$-c$ 2d CFT on $S^1$ set up by inserting an operator at the origin of the Euclidean disk.

We observe that \eqref{rc} has some nice structural properties:
\begin{enumerate}
\item[(1a)] $S_{IR}(\vec{P})$, \eqref{sbulk}, is itself the von Neumann entropy of a  state $\rho^{IR}_{\vec{P}}$ that one can explicitly construct up to a choice of base point along a Euclidean cycle in Virasoro TQFT \cite{Collier:2023fwi, Collier:2024mgv}.	
\item[(1b)] 
	Under a standard assumption, \footnote{Specifically, below we will be able to show \eqref{e15} explicitly, i.e. by direct evaluation of $\log S_{P_i 1}$ on $\vec{P}^*$, for a single interval in the vacuum state and in states set up by inserting a light operator at the origin of the Euclidean disk. We will not be able to do this for multiple intervals or in states set up with heavy operator insertions because in those cases, we don't know the conformal blocks appearing in $p_{\vec P}$ in closed form. However, in those cases we can argue indirectly for \eqref{e15} under the assumption flagged in footnote 1, to the same level of rigor as other recent works \cite{Geng:2025efs, Bao:2025plr} in the literature.}  
	the $``\log S_{P1}"$ part of \eqref{rc}, evaluated at the peak of the distribution $p_{\vec P}$, dominates the entanglement entropy to leading order in $c$, 
\be\label{e15}
S(A, |\psi\rangle) \simeq \left. \sum_i \log S_{P_i 1}\right|_{\vec{P} = \vec{P}^*}.
\ee
\end{enumerate}

\item {\bf Boundary interpretation.} We then argue that \eqref{rc} has the {interpretative} properties that 

\begin{enumerate}
\item[(2a)] It structurally resembles an algebraic entanglement entropy  
	for a subalgebra $\AAA_{IR}$ with a center labeled by $\vec{P}$, and with $\sum_i \log S_{P_i 1}$ playing the role of a sector multiplicity.
\item[(2b)] If indeed there is such an underlying algebra, ((1a) suggests that) a natural candidate for $\AAA_{IR}$ is the Virasoro algebra of the CFT on region $A$.
\item[(2c)] If indeed (2b) is correct, (the fact that on the one hand physical CFTs consist of Virasoro modules with no intrinsic multiplicity, and on the other hand $S_{P 1}$ coincides with the Cardy density of heavy primaries in a holographic CFT, suggests that) the $\sum_i\log S_{P_i1}$ term admits a natural interpretation as the density-of-states contribution to a sector multiplicity after heavy primaries of the BCFT in region $A$ are coarse-grained into Liouville momentum windows centered at the $P_i$'s.
\item[(2d)] Our explicit 
construction of the $p_{\vec P}$'s appearing in \eqref{rc} provides quantitative criteria for the amount of coarse-graining necessary and/or allowed for the consistency of the rest of the interpretation.
\end{enumerate}

\item {\bf Large-$c$ holographic identification.} At the same time, \eqref{rc} has the same leading large-$c$ organization as the bulk generalized entropy. We are thus led to the holographic identification that whenever \eqref{e15}  is true, 
	the Ryu-Takayanagi area term in \eqref{rt} is dual to the $``\log S_{P1}$" term in \eqref{rc}, evaluated at the peak of $p_{\vec P}$: 

\be\label{main}
\left.\frac{\mbox{Area}(\gamma_A)}{4G_N} = \sum_i \log S_{P_i1}\right|_{\vec{P}=\vec{P}^*}\,.
\ee

Conditioned on our boundary interpretation (point 2), the RT area can therefore be identified with the density-of-states part of the dominant sector multiplicity after coarse-graining heavy Virasoro primaries in the BCFT on region $A$. 

\item {\bf Identification of prospective ``hardware" carrying the RT entropy.} This result points to a natural set of candidate physical carriers of the RT entropy. 
\begin{enumerate}
\item[(4a)] From a boundary point of view, \eqref{main} suggests that the RT area should be attributed to an entropic contribution of Virasoro intertwiners, i.e. bilocal operators between the BCFT's on the regulated regions $A$ and $\bar A$ built from Virasoro-charged operators in each region whose representation labels are compatible with the Virasoro charge of the global state, in a coarse-grained neighborhood of the dominant $P$-value. \footnote{Refining an earlier conjecture of \cite{Casini:2019kex}.}
\item[(4b)] From a ``pre-geometric" bulk point of view, \eqref{main} suggests that the RT area may be an edge term for a line operator at the dominant representation $P^*$ in a non-compact Chern-Simons-like description of the bulk, or in other words that the bulk duals of the intertwiners in item (4a) are such line operators. This follows from the fact that $S_{P1}$ \eqref{sp1} also coincides with a Plancherel measure for a quantum deformation of $sl(2)$ \cite{Ponsot:1999uf, Blommaert:2018iqz}.
\end{enumerate}
\end{enumerate}

\noindent The rest of this paper is organized as follows.

\noindent In section {\bf \ref{s2}}, we derive \eqref{rc} in the situations mentioned above, and explain the structural properties.

\noindent In section {\bf \ref{s3}}, we explain the interpretation in the boundary CFT.

\noindent In section {\bf \ref{s4}}, we present the identification \eqref{main} and discuss  implications for AdS$_3$/CFT$_2$ holography.

\noindent Finally, we briefly discuss some directions for future work in section {\bf \ref{s5}}.

\section{Cross-channel entropy formula in holographic 2d CFTs}\label{s2}

We first show that \eqref{rc} is true across a selection of states and subregions in a holographic 2d CFT.

\subsection{A single interval in the vacuum state}\label{s21}

We start with the warm-up case of a single interval $A=[0,L]$ in the vacuum state of a 2d CFT on $S^1$. This case was previously discussed in \cite{Wong:2022eiu, Lin:2021veu}.

\begin{figure}
	\centering
	\includegraphics[height=1.2in]{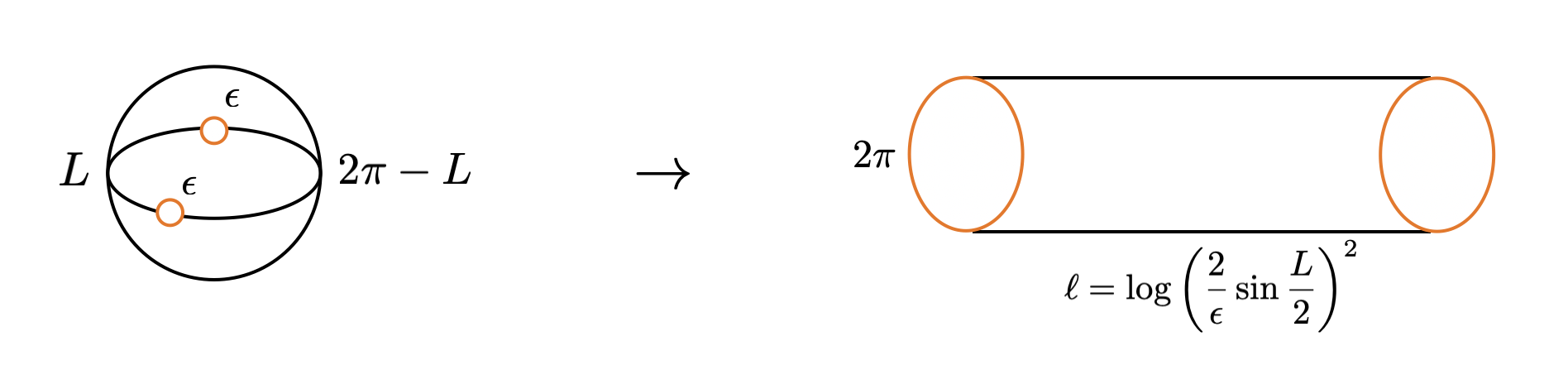}
	\caption{Preparing the factorized vacuum state on $S^1$ and the conformal map to a cylinder.}
	\label{fig1} 
	\end{figure}

The vacuum state of a CFT on $S^1$ can be set up by doing a Euclidean path integral on a disk ending on the $S^1$. To properly define an entanglement calculation, we must split the $S^1$ into the region $A$ and its complement. To prepare this factorized state with a Euclidean path integral, we excise a region of size $\varepsilon$ around the entangling boundary $\partial A$, here consisting of the two endpoints of $A$, at which we put boundary conditions $|a_1\rangle$, $|a_2\rangle$ \cite{Ohmori:2014eia}. This causes the $n$-replicated surface to have the topology of a cylinder. After a conformal transformation, we can take this cylinder to have circumference $2\pi n$ and length 
\be\label{lengthl}
\ell =  \log \left(\frac 2\varepsilon \sin \frac L 2 \right)^2\,.
\ee
See Fig. \ref{fig1}. See Section 3.1 of \cite{Cardy:2016fqc} for the exact conformal map.

The Euclidean partition function of the CFT on the $n$-replicated cylinder can be computed in a few different ways. In the ``closed string channel", we view it as an amplitude between the boundary states,
\be\label{e22}
Z_n = \langle a_1|\exp\left(\frac \ell n (\frac{c+\bar c}{24}- L_0 - \bar L_0) \right) |a_2\rangle = \langle a_1 |\tilde q(n)^{\frac 12 (L_0 + \bar L_0 - \frac{c+\bar c}{24})}|a_2\rangle\,,
\ee
where $\tilde q(n) = e^{-2\ell/n}$. As long as the CFT has a gap, any choice of the $|a_i\rangle$'s that overlaps with the vacuum Ishibashi state $|0\rangle\rangle$ will give the same answer as replacing them with $|0\rangle\rangle$, 
to leading order in $\varepsilon$ \cite{Wong:2022eiu}. 
This lets us approximate 
\be\label{sicc}
Z_n \sim \langle\langle 0| \tilde q(n)^{\frac 12(L_0 + \bar L_0 - \frac{c+\bar c}{24})} |0\rangle\rangle  = \chi_0(\tilde q(n))\,,
\ee 
where
\be
\chi_0(q) = \frac{(1 -  q) q^{-\frac{c-1}{24}}}{\eta( q)}
\ee
is the vacuum character on the torus with modular parameter $q$, and $\eta(q) = q^{\frac 1 {24}}\prod_{n=1}^\infty (1-q^n)$ is the Dedekind eta function. For large  $\ell$, we can further approximate \eqref{sicc} as 
\be\label{n25}
Z_n \sim e^{\frac{c\ell}{12n}}\,.
\ee
 Plugging \eqref{n25} into the replica trick formula,
 \be\label{replica}
 S = -\frac{\partial}{\partial n}\log \left. \left(\frac{Z_n}{(Z_1)^n} \right)\right|_{n = 1} = -\frac{\left.\partial_n Z_n\right|_{n \rightarrow 1}}{Z_1} + \log Z_1\,,
 \ee
 we find the standard universal answer for the entanglement entropy of a 2d CFT across a single interval: 
 \be\label{sinte}
 S(A) \sim \frac {c \ell}{6}\,.
 \ee
 
 However, although \eqref{sicc} is exact to leading order in $1/\varepsilon$, its dependence on the replica index $n$ is not conducive to putting the entropy in the form of \eqref{rc}. To put the entropy in the form of \eqref{rc}, we apply crossing symmetry. In this case, applying a single modular transformation to the vacuum block $\chi_0(\tilde q(n))$ in \eqref{sicc}, we find  
\be\label{s24p}
Z_n \sim \int_0^\infty dP\, S_{P1}\, \chi_P(q(n))\,,
\ee
where 
\be
\chi_P(q) = \frac{q^{P^2}}{\eta(q)}
\ee
is a Virasoro character in a non-vacuum representation, and $q(n) = e^{-{2\pi^2 n}/{\ell}}$\,.
In terms of the normalized integrand of $Z_1$ in \eqref{s24p}, 
\be\label{pp}
p_P= Z_1^{-1} S_{P1}\, \chi_P(q(1))\,,
\ee
we can exactly rewrite \eqref{s24p} as
 \be\label{e25}
 Z_n \sim Z_1^n \int_0^\infty dP\, (p_P)^n (S_{P1})^{1-n} \frac{\chi_P(q(n))}{\chi_P(q(1))^n}\,.
 \ee
Plugging \eqref{e25} into \eqref{replica} gives
 \be\label{e27}
 S = \int dP\, p_P(-\log p_P + \log S_{P1} - \left(\frac{\left.\partial_n \chi_P(q(n))\right|_{n \rightarrow 1}}{\chi_P(q(1))} - \log \chi_P(q(1)) \right))\,.
 \ee 
This has the advertised form of \eqref{rc}.

Note that the results in this section did not depend on the CFT having a semiclassical bulk dual. Rather, ``vacuum block dominance" in \eqref{sicc} follows from the smallness of the geometric regulator $\varepsilon$. The situation will be different in the multiple-interval and excited state cases. 

\subsection{Two intervals in the vacuum state} \label{s22}

Next, we discuss the case of two disjoint intervals, $A = A_1 \cup A_2$ for $A_1 = [0, L_1]$ and $A_2 = [x, x+L_2]$, in the vacuum state of the 1+1$d$ CFT. 

\begin{figure}
	\centering
	\includegraphics[height=1.5in]{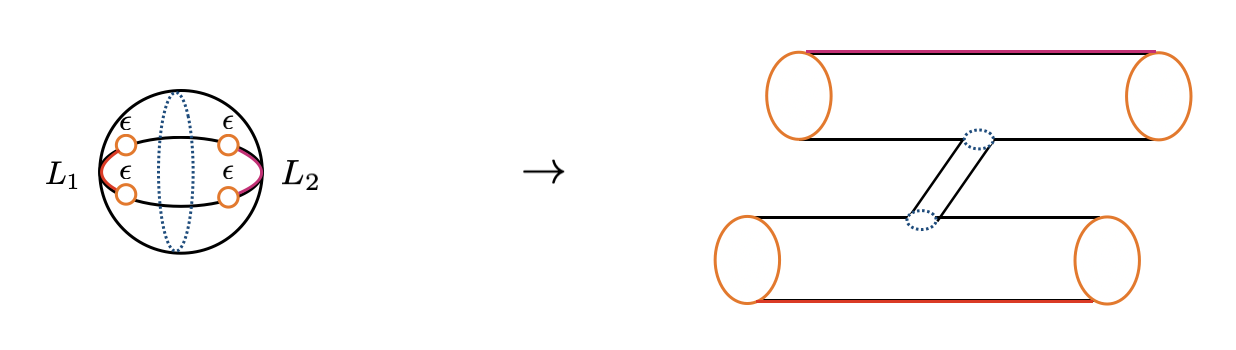}
	\caption{After regularization, the replica partition function for the two-interval case with replica index $n=1$ lives on a space with the topology of two cylinders joined by a tube. This figure is meant to illustrate the topology only and is not drawn to scale. For $Z_n$, the manifold would have $n$ connecting tubes spaced along the short cycle at the midpoint of the long direction of the cylinders.} 
	\label{fig4} 
	\end{figure}

In this case, we start again with the global vacuum state of the CFT on $S^1$ set up by doing a Euclidean path integral on a disk. To properly define the entanglement calculation, we cut out regions of size $\varepsilon$ around each of the four endpoints of $A_1 \cup A_2$, putting boundary conditions $|a_1\rangle$, $\dots$, $|a_4\rangle$ there. This causes the manifold on which we compute the $n$-replicated partition function $Z_n$ to have the topology of a pair of cylinders joined along $n$ tubes. 
	See Fig. \ref{fig4} for an illustration of the situation for $n=1$. See \cite{Calabrese:2010he, Coser:2015dvp} for earlier discussions of this replica surface prior to regularizing the entanglement calculation.

At large $c$ and assuming a sparse light spectrum, it's expected that the partition function $Z_n$ will be well-approximated by the vacuum block in some channel \cite{Hartman:2013mia}. For small $\varepsilon$, the vacuum block will dominate in the channel where we take whichever pairs of endpoints/boundaries are closest along the $S^1$ to be the ends of the cylinders in Fig. \ref{fig4}, let the identity propagate between them, and also let the identity propagate in the tubes that join the cylinders. This generalizes the ``closed string channel" of section \ref{s21}. There are two ways to pair the boundaries: with their nearest neighbor to either the right or left along the $S^1$. Let us call these possibilities the ``S" and ``T" pairings. Without loss of generality, let us first analyze the case where the vacuum block dominates in the pairing labeled by ``S".

\begin{figure}
	\centering
	\includegraphics[height=1in]{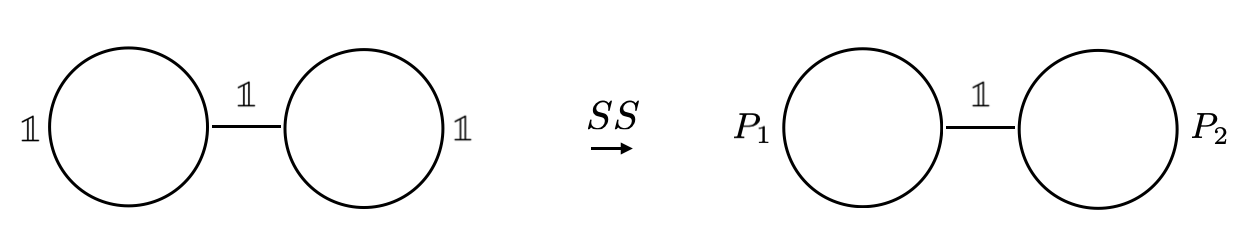}
	\caption{$S$-moves that we use to go between the channel where the vacuum block dominates and the one used to compute the cross-channel entropy, for the two-interval case. On the left side, the circles are the long directions of each of two cylinders ($<>$ torus after doubling), and on the right side the circles are the short cycles of each cylinder. Before applying these S-moves, we collapse the $n$ identity tubes joining the cylinders to one via trivial $F$-moves, and afterwards we undo the collapse, so the full sequence of moves is $F \rightarrow S \rightarrow F^{-1}$.} 
	\label{fig5} 
	\end{figure}

To put the entanglement entropy in the form \eqref{rc}, we apply the minimal set of crossing moves that amount to doing an independent $S$-move on each cylinder/pair of  entangling boundaries. We first fuse the $n$ identity tubes using trivial $F$-moves (with identity intermediate states) to collapse the graph to a pair of cylinders that are each attached to a single vacuum tube. We then apply a modular $S$-move on (the doubled torus that corresponds to) 
	each cylinder. (See Figure \ref{fig5}.) Finally, we reverse the $F$-moves that we did at the start. 

This set of crossing moves yields
\be\label{e210}
Z_n \sim \mathcal{F}_{0}^{closed (S)}(n) = \int dP_1\, S_{P_1 1} \int dP_2\, S_{P_2 1}\, \mathcal{F}^{L(S)}_{P_1P_2}(n)\,,
\ee
where the $(S)$ index denotes the choice of how to pair the endpoints; the ``closed" channel where the vacuum block dominates is as described above; and $\mathcal{F}^{L(S)}_{P_1P_2}(n)$ is the ``ladder" block with momenta $P_1$ and $P_2$ propagating in the ``open string" directions of each of the two cylinders respectively, and the identity propagating along the $n$ tubes that join them. 

In terms of the normalized integrand of $Z_1$ in \eqref{e210}, 
	\be
	p_{P_1P_2}^{(S)} = Z_1^{-1}S_{P_1 1} S_{P_2 1}\, \mathcal{F}^{L(S)}_{P_1P_2}(n=1)\,,
	\ee 
we can exactly rewrite \eqref{e210} as 
	\be\label{e212}
	Z_n \sim Z_1^n\int dP_1 \int dP_2\, (p_{P_1P_2}^{(S)})^n (S_{P_1 1} \,S_{P_2 1})^{1-n}\, \frac{\mathcal{F}^{L(S)}_{P_1P_2}(n)}{\mathcal{F}^{L(S)}_{P_1P_2}(1)^n}\,.
	\ee
Plugging \eqref{e212} into the replica trick formula \eqref{replica} yields 
	\be\label{e213}
	S = \int dP_1\, \int dP_2\, p_{P_1P_2}^{(S)}(-\log p_{P_1 P_2}^{(S)} + \log S_{P_1 1} + \log S_{P_2 1} - \left(\frac{\left.\partial_n \mathcal{F}^{L(S)}_{P_1P_2}(n)\right|_{n\rightarrow 1}}{\mathcal{F}^{L(S)}_{P_1P_2}(1)} - \log \mathcal{F}^{L(S)}_{P_1P_2}(1) \right))
	\ee 
which has the promised form of \eqref{rc}.

Eq. \eqref{e213} is exact up to assuming that the $n$-replicated partition function is dominated by the vacuum block in the ``S" way of pairing the 
replica manifold boundaries that one gets by regularizing the entangling region. Had the vacuum block dominated in the ``T" way of pairing the boundaries, the same logic would yield \eqref{e213} with the label $(S)$ swapped for $(T)$. To collect the two possibilities, we write 
\be\label{2intf}
S = \min_{S,T} \int d P_1\, \int d P_2\, p_{P_1P_2}^{(S/T)} (-\log p_{P_1P_2}^{(S/T)} + \log S_{P_1 1} + \log S_{P_2 1} - \left(\frac{\left.\partial_n \mathcal{F}^{L(S/T)}_{P_1P_2}(n)\right|_{n\rightarrow 1}}{\mathcal{F}^{L(S/T)}_{P_1P_2}(1)} - \log \mathcal{F}^{L(S/T)}_{P_1P_2}(1) \right))
\ee
for 
\be\label{pp1p2}
p_{P_1P_2}^{(S/T)} = Z_1^{-1} S_{P_1 1}S_{P_2 1}\mathcal{F}^{L(S/T)}_{P_1P_2}(1)\,,
\ee
minimizing over choices of ``ladder" block
where we decompose the Euclidean manifold that sets up the entanglement calculation into two cylinders with nontrivial momenta propagating in their ``open string" directions, and the identity propagating in the tubes that join them. 

Comments:
\begin{itemize}
\item[(*)] In this case, we had to make the large$-c$/sparse spectrum assumption of \cite{Hartman:2013mia} to invoke vacuum block dominance, unlike the single-interval case where we got it for free. Once we make this assumption, all of the terms in \eqref{2intf} are universal in the sense of \cite{Hartman:2013mia} (i.e., they depend on $c$ and the geometric moduli, but not the OPE coefficients or fine spectrum of the CFT).

\item[(*)] In eqs. \eqref{e210} - \eqref{2intf}, $\mathcal{F}^{L(S/T)}_{P_1P_2}(n)$ is not a priori well-defined for non-integer $n$. 
However, we can use the usual twist operator construction to define it at non-integer $n$. I.e., 
we can 
replace the $n$-sheeted replica geometry with the $n=1$ geometry with two twist operators of weights $h =\bar h = \frac c{24}(n-\frac 1n)$ on each cylinder, one each at the end of each cylinder at the same replica angle (i.e. at the ends of the pink lines in Figure \ref{fig4}), then let the Virasoro block with momenta  $P_1$ and $P_2$ around the closed cycles of the cylinders on the $n=1$ geometry with twist operator insertions define the analytic continuation. Note that we only use the twist operator construction to define $\mathcal{F}^{L(S/T)}_{P_1P_2}(n)$ as a sub-step in the above manipulations, rather than using it to do the full entanglement calculation.
\item[(*)] 	A similar formula to \eqref{e210} was recently derived in \cite{Geng:2025efs} from an independent logical chain. Those authors obtained a similar formula by first triangulating the Euclidean replica manifold to represent $Z_n$ as a BCFT tensor network with OPE coefficients assigned to vertices, then using the ETH to contract pairs of OPE coefficients, and finally using crossing moves. (See \cite{Bao:2025plr} for more on their method.) We instead reach the same type of formula by assuming vacuum block dominance. We will refer again to their work in section \ref{s25} when we discuss the saddle-point analysis of \eqref{2intf}.
\end{itemize}

\subsection{A single interval in an excited state}\label{s23}

The entanglement entropy across a single interval $A = [0, L]$ in an excited state of a 2d CFT on $S^1$, set up by inserting an operator $\OO$ of dimension $h_\OO = \bar h_\OO$ at the origin of the Euclidean disk, can be organized in much the same way.

\begin{figure}
	\centering
	\includegraphics[height=1.2in]{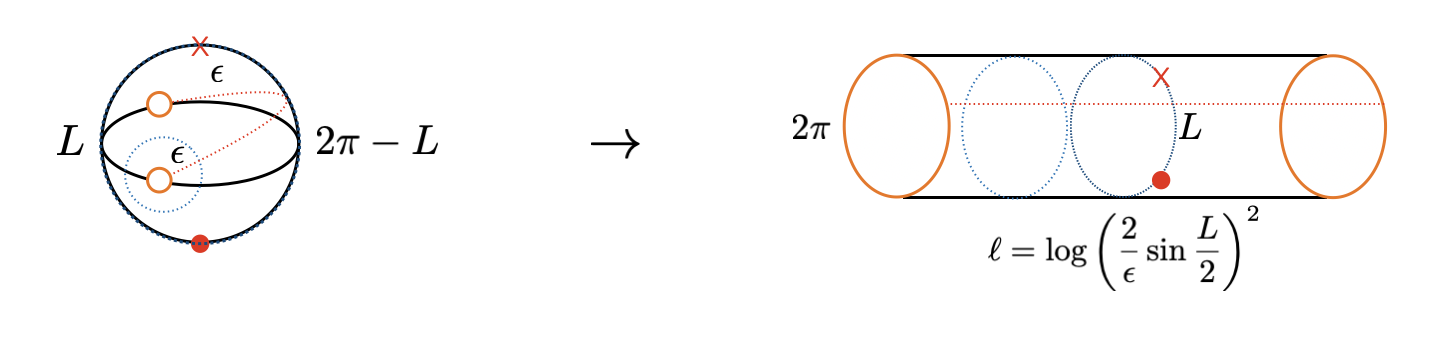}
	\caption{Preparing the factorized excited state on $S^1$ and the conformal map to a cylinder.}
	\label{fig2} 
	\end{figure}

In this case, the $n$-replicated manifold is the same as in the single-interval vacuum case (section \ref{s21}) up to the insertion of $2n$ copies of $\OO$ on the replica surface. Under the same conformal map as in that case, it becomes a cylinder of length $\ell$, \eqref{lengthl}, and circumference $2\pi n$, with the operator insertions spaced alternating distances of $L$ and $2\pi -L$ around the closed cycle of the cylinder at the midpoint of the long direction. See Fig. \ref{fig2} for an illustration of the situation for $n=1$.

At large $c$ and assuming a sparse light spectrum, the vacuum block will dominate in the channel where we first contract nearest-neighbor operators to the identity, then let the identity propagate along the long (``closed string") direction of the cylinder \cite{Hartman:2013mia}. Depending on whether $L$ is less than or greater than $2\pi-L$, we contract a given operator with the one to its left or right. Let us call these possibilities the ``S" and ``T" pairings, and analyze WLOG the ``S" case where $L < 2\pi -L$.

\begin{figure}
	\centering
	\includegraphics[height=1.8in]{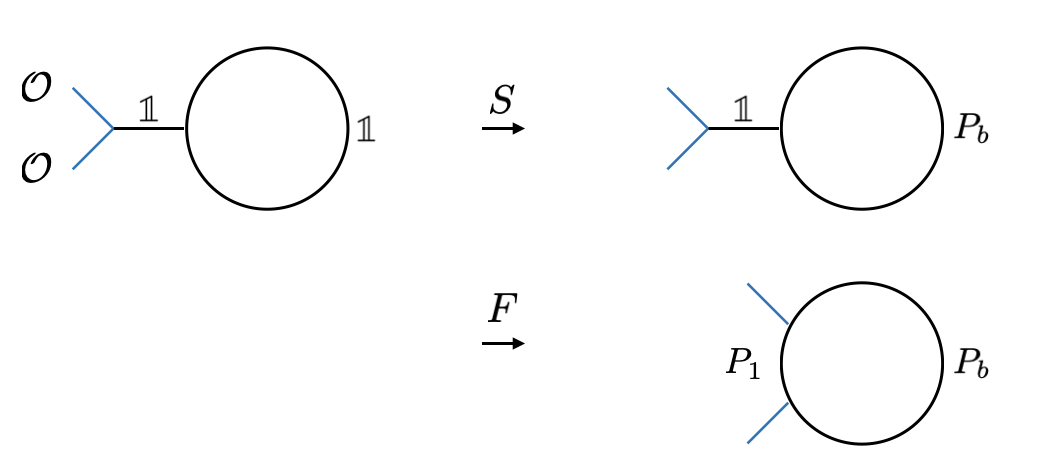}
	\caption{$S$- and $F$- moves that we use to go between the channel where the vacuum block dominates and the one used to compute the cross-channel entropy, for the excited state case at replica index $n=1$. On the left side, the circle is the long direction of the cylinder ($<>$ torus after doubling), and on the right side the circle is the short cycle of the cylinder. We also perform the reverse set of $F$-moves at the start before applying the modular S-move (not shown), so the full sequence of moves is $F \rightarrow S \rightarrow F^{-1}$.}
	\label{fig3} 
	\end{figure}

To put the entanglement entropy in the form \eqref{rc}, we again apply crossing moves that amount to isolating an $S$-move on the cylinder that joins the pair of replica manifold boundaries. We first fuse nearest-neighbor pairs of external operators to the identity, and use trivial $F$-moves to collapse the identity legs to a single insertion of the identity on the cylinder. We then apply a modular $S$-move on the (doubled torus that corresponds to the) cylinder, and finally reverse the $F$-moves that we did at the start. See Fig. \ref{fig3} for an illustration of the situation for $n=1$.

This set of crossing moves yields
\be\label{znexc}
	Z_n \sim \mathcal{F}^{closed (S)}_{0}(n) = \int dP_b\, S_{P_b 1}  \int dP_1 \mathbb{F}_{P_1 1}\,\left[\begin{tabular}{cc} $P_\OO$ & $P_b$ \\ $P_\OO$ & $P_b$ \end{tabular} \right] \dots \int\, dP_n\, \mathbb{F}_{P_n 1}\left[\begin{tabular}{cc} $P_\OO$ & $P_b$ \\ $P_\OO$ & $P_b$ \end{tabular} \right] \mathcal{F}^{N(S)}_{P_1, P_b, \dots, P_n, P_b}(n)\,,
\ee
where the $(S)$ index denotes the choice of how to pair the operator insertions; the ``closed" channel where the vacuum block dominates is as described above; and $\mathcal{F}^{N(S)}_{P_1, P_b, \dots, P_n, P_b}(n)$ is the ``necklace" channel block with momenta $P_1$, $P_b$, $\dots$, $P_n$, $P_b$ propagating along alternating segments of length $L$ and $2\pi -L$ between operator insertions in the ``open string" direction. Moreover, $\mathbb{F}$ is the fusion kernel
\be\label{fusion}
\mathbb{F}_{P_i 1}\left[\begin{tabular}{cc} $P_\OO$ & $P_b$ \\ $P_\OO$ & $P_b$ \end{tabular} \right] = S_{P_i 1}\, C_0(P_i, P_\OO, P_b)\,,
\ee
where 
\be\label{c0}
	C_0(P_1, P_2, P_3) = \frac{1}{\sqrt{2}}\frac{\Gamma_b(2Q)}{\Gamma_b(Q)^3}\frac{\prod_{\pm\pm\pm}\Gamma_b(\frac Q 2 \pm iP_1 \pm iP_2 \pm iP_3)}{\prod_{k=1}^3\Gamma_b(Q+ 2iP_k)\Gamma_b(Q - 2iP_k)};
\ee
$Q$ is defined via $c = 1+6Q^2$; $\Gamma_b$ is the Barnes double gamma function; and $S_{P1}$ is the modular S-matrix element \eqref{sp1} that we encountered before.

Assuming that $Z_n$ is peaked around configurations with replica symmetry, \footnote{ Note this is an extra assumption that we make beyond either of the previous examples.} we can further simplify \eqref{znexc} to 
\be\label{e220}
Z_n \sim \int dP_b\, S_{P_b 1} \int dP_i\,  \mathbb{F}_{P_i 1}\left[\begin{tabular}{cc} $P_\OO$ & $P_b$ \\ $P_\OO$ & $P_b$ \end{tabular} \right]^{n} \mathcal{F}^{N(S)}_{P_i, P_b, \dots P_i, P_b}(n)\,.
\ee 
In terms of the normalized integrand of $Z_1$ in \eqref{e220}, 
\be
p_{P_iP_b}^{(S)} = Z_1^{-1}S_{P_b 1}\, \mathbb{F}_{P_i 1}\left[\begin{tabular}{cc} $P_\OO$ & $P_b$ \\ $P_\OO$ & $P_b$ \end{tabular} \right] \mathcal{F}^{N(S)}_{P_i, P_b}(1)\,,
\ee
we can rewrite \eqref{e220} as
\be\label{e20}
Z_n \sim Z_1^n \int dP_b \int dP_i \, (p_{P_iP_b}^{(S)})^n (S_{P_b 1})^{1-n} \frac{\mathcal{F}^{N(S)}_{P_i, P_b, \dots P_i, P_b}(n)}{\mathcal{F}^{N(S)}_{P_i, P_b}(1)^n} \,.
\ee
Plugging \eqref{e20} into the replica trick formula \eqref{replica} yields
\footnote{
As before, we can use a twist operator construction to formally define $\mathcal{F}^{N(S)}_{P_i, P_b, \dots P_i, P_b}(n)$ around $n=1$.
}
\be\label{e222}
S = \int dP_b \int dP_i\, p_{P_iP_b}^{(S)}(-\log p_{P_iP_b}^{(S)} + \log S_{P_b 1} - \left(\frac{\left.\partial_n\mathcal{F}^{N(S)}_{P_i,P_b, \dots P_i, P_b}(n)\right|_{n\rightarrow 1}}{\mathcal{F}^{N(S)}_{P_i, P_b}(1)} - \log \mathcal{F}^{N(S)}_{P_iP_b}(1)\right))\,,
\ee
which has the desired form of \eqref{rc}. 

Had the vacuum block dominated in the ``T" way of pairing the external operator insertions, the same logic would yield \eqref{e222} with the label $(S)$ swapped for $(T)$. To collect the two possibilities, we write 
\be\label{eans}
S = \min_{S,T} \int dP_b \int dP_i\, p_{P_iP_b}^{(S/T)}(-\log p_{P_iP_b}^{(S/T)} + \log S_{P_b 1} - \left(\frac{\left.\partial_n\mathcal{F}^{N(S/T)}_{P_i,P_b, \dots P_i, P_b}(n)\right|_{n\rightarrow 1}}{\mathcal{F}^{N(S/T)}_{P_i, P_b}(1)} - \log \mathcal{F}^{N(S/T)}_{P_iP_b}(1)\right))\,,
\ee
for
\be\label{ppipb}
p_{P_iP_b}^{(S/T)} = Z_1^{-1}S_{P_b 1}\, \mathbb{F}_{P_i 1}\left[\begin{tabular}{cc} $P_\OO$ & $P_b$ \\ $P_\OO$ & $P_b$ \end{tabular} \right] \mathcal{F}^{N(S/T)}_{P_i, P_b}(1)\,.
\ee

\subsection{The $S_{IR}(\vec P)$ term is the von Neumann entropy of a state in Virasoro TQFT}\label{s24}

Having derived \eqref{rc} for each case that we wish to consider, we next discuss its structural properties. The first of these is that the $S_{IR}(\vec{P})$ term in \eqref{rc} can be understood as the von Neumann entropy of a state in Virasoro TQFT, which is a bulk theory of pure $AdS_3$ gravity that keeps only the universal Virasoro/stress tensor part of a would-be boundary CFT \cite{Collier:2023fwi, Collier:2024mgv}.

Across the examples in sections \ref{s21} - \ref{s23}, $S_{IR}(\vec P)$ had the form of an application of the replica trick \eqref{replica} to a family of conformal blocks $\mathcal{F}_{\vec P}(n)$ on the $n$-replicated manifold for each full entanglement calculation:

\be\label{sbulk2}
S_{IR}(\vec P) = -\frac{\partial_n\mathcal{F}_{\vec P}(n)|_{n \rightarrow1}}{\mathcal{F}_{\vec P}(1)} + \log \mathcal{F}_{\vec P}(1)\,.
\ee

On the other hand, according to Virasoro TQFT, a conformal block on a Riemann surface 
	labeled by momenta $P_i$ propagating along select cycles equals the Virasoro TQFT partition function on the three-manifold where we make those cycles non-contractible and wrap them with Wilson lines labeled by the $P_i$'s. Applied case-by-case to the situations discussed in section \ref{s2},

\begin{itemize}
\item[(*)] For a single interval in the vacuum state, the block $\mathcal{F}_{\vec P}(n)$ is the Virasoro character on the torus that we get 	after doubling the cylinder in Fig. \ref{fig1}, in the channel where we take space to consist of the entangling interval and its doubled copy, and let momentum $P$ propagate in the other (replica) direction. Under Virasoro TQFT,  $\chi_{P}(q(n))$ is the Virasoro TQFT partition function on the solid torus where we fill in the spatial disk and wrap the (now non-contractible) replica cycle with a Wilson line labeled by $P$. See Fig. \ref{fig6}. 
	By running the usual replica trick logic in reverse, we conclude that $\rho_P^{IR}$ is the thermal state in Virasoro TQFT on the disk with a primary $P$ at the origin, that one gets from cutting the path integral interpretation for $\chi_P(q(n=1))$ along a slice of constant Euclidean (replica) time.
	
	\begin{figure}
	\centering
	\includegraphics[height=1.8in]{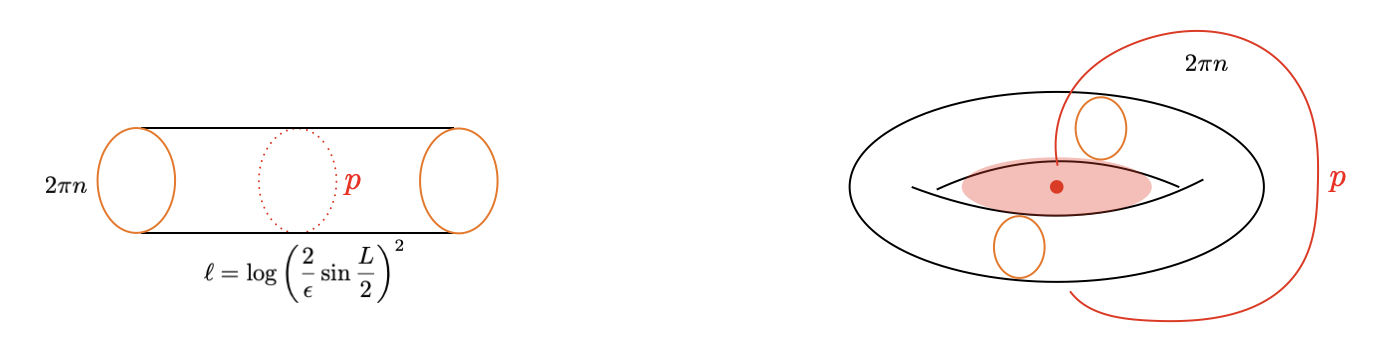}
	\caption{For a single interval in the vacuum state of a CFT, the character $\chi_p(q(n))$ associated with Virasoro primary $P$ propagating around the ``open string" channel of the replica manifold (left) equals the Virasoro TQFT partition function on the solid torus where we fill in the dual cycle 
		and wrap the now non-contractible replica cycle with a Wilson line labeled by the same $P$ (right). 
		}
	\label{fig6} 
	\end{figure}

\item[(*)] For two intervals in the vacuum state of the CFT, the block $\mathcal{F}_{\vec P}(n) = \mathcal{F}^{L(S/T)}_{P_1P_2}(n)$ is the one with momenta $P_1$, $P_2$ propagating around the open string (replica) directions of each of the two cylinders of the manifold described in section \ref{s22}. Under Virasoro TQFT, this block equals the Virasoro TQFT partition function on the bulk three-manifold where after doubling each cylinder to get a torus, we fill in each disk bounded by the image of an entangling interval and its copy, and wrap the other (now non-contractible) cycle with a Wilson line labeled by $P_1$ or $P_2$; basically two copies of the construction depicted in Fig. \ref{fig6}. Running the usual replica trick logic in reverse, 
	we conclude that $\rho^{IR}_{P_1P_2}$ is in this case a nearly-thermal state in Virasoro TQFT on the product of two solid disks with primaries $P_1$, $P_2$ at their origins, that one gets from cutting the path integral interpretation for $\mathcal{F}^{L(S/T)}_{P_1P_2}(n)$ along a slice of constant Euclidean (replica) time.

	Note that in this case the replica cycle that we cut along is not $U(1)$-symmetric (since the $U(1)$ symmetry is broken by the connecting tube in Fig. \ref{fig4}), and different choices of the replica angle to cut along give different density operators with the same  $\Tr(\rho^{IR}_{\vec P})^n$'s. The construction described here defines $\rho^{IR}_{P_1P_2}$ up to this ambiguity.

\item[(*)] For a single interval in an excited state of the CFT set up by inserting an operator at the origin of the Euclidean disk, $\mathcal{F}_{\vec P}(n) = \mathcal{F}^{N(S/T)}_{P_iP_b}$ is the ``necklace" block defined in section \ref{s23}. In this case, the Virasoro TQFT dual after doubling lives on a solid torus whose core carries a piecewise Wilson line network around the replica cycle with $n$ alternating segments labeled by $P_i$, $P_b$, anchored to boundary Wilson lines that carry the $\OO$-insertions. Cutting the solid torus at a replica angle that goes through the $P_b$ arc, we conclude that $\rho^{IR}_{PiP_b}$ is in this case a nearly thermal state in Virasoro TQFT on a solid disk with a primary $P_b$ at the origin, up to the Wilson line network structure in the Euclidean path integral that sets up the state. 

	Again, since in this case the replica cycle is not $U(1)$-symmetric, cuts at different replica angles give different density operators with the same R\'enyi spectrum. The construction described here defines $\rho_{P_iP_b}^{IR}$ up to this ambiguity.
	In particular, we can choose to cut through the $P_b$ segment as a convention to match the $\log S_{P_b1}$ factor in \eqref{eans}.
\end{itemize}

\subsection{The $`\log S_{P1}$' term evaluated at the peak of $p_{\vec P}$ dominates at large $c$}\label{s25}

A second structural property is that across our examples, under assumptions made elsewhere in the recent literature, the $\log S_{P1}$ term evaluated at the peak of the distribution $p_{\vec P}$ dominates \eqref{rc} to leading order in $c$. I.e., 
\be\label{sp1p}
S(A) \simeq \left. \sum_i \log S_{P_i 1}\right|_{\vec{P} = \vec{P}^*}\,,
\ee
for $\vec{P}^*$ the peak of each distribution $p_{\vec P}$. 
To show this, we will proceed case-by-case through the situations considered in sections \ref{s21} - \ref{s23}.

\subsubsection{Single-interval vacuum case}

In the single-interval vacuum case, we can show \eqref{sp1p} quite explicitly. In this case, the analog of $p_{\vec P}$ is  
  \be\label{pp2}
p_P= Z_1^{-1} S_{P1}\, \chi_P(q(1))\,,
\ee
where $\chi_P(q(1))$ is a Virasoro character on the torus with $q = e^{\frac {-2\pi^2}{\ell}}$. See \eqref{pp} and \eqref{e27}. When $c \gg 1$, $p_P$ is dominated by a peak at 
\be\label{pstar}
P_* = \frac{b^{-1}\ell}{2\pi}\,.
\ee
Evaluating $\log S_{P1}$ on $P_*$ recovers the full $O(c)$ part of the entanglement entropy, \eqref{sinte}:
\be
\left. \log S_{P1}\right|_{P = P_*} \approx 2\pi b^{-1} P_* = \frac{c\ell}{6}\,.
\ee 
Moreover, the saddle-point values of the Shannon and bulk terms in \eqref{e27} contribute only at $O(c^0)$, while corrections to the Laplace expansion are non-perturbatively small in $\ell$ and $c$. \footnote{ In particular, we can read off from $p_P$ approaching a Gaussian at large $\ell$ and $c$ that the Shannon term contributes $\frac 12 \log \ell$ at $O(c^0)$.}

\subsubsection{Two-interval vacuum case}\label{s252}

For the two-interval case, we cannot currently show \eqref{sp1p} by direct discovery of $\vec{P}^*$ followed by evaluating $\sum_i \log S_{P_i 1}$ on it. This is because $p_{\vec P}$ in this case, \eqref{pp1p2}, contains the ``ladder" Virasoro block discussed in section \ref{s22}, which is not currently known in closed form. 
However, we can indirectly argue that non-``$\log S_{P1}$"- type terms in \eqref{rc} are $O(c^0)$ using an argument of, and assumption from, \cite{Geng:2025efs, Bao:2025plr}.

Ref. \cite{Bao:2025plr} first observes that as long as the integral form of $Z_n$, \eqref{e210}, is controlled by a saddle near $n=1$, a sufficient condition for \eqref{sp1p} is for the conformal block appearing therein to satisfy \footnote{See section 4.1.2 of \cite{Bao:2025plr}.}
\be\label{flin}
\mathcal{F}_{\vec P}(n) = \mathcal{F}_{\vec P}(1)^n
\ee
to leading order in $c$.
Abstractly, consider a family of $Z_n$'s with the form
\be\label{defzn}
Z_n = \int \prod_i dP_i\, S_{P_i 1}\mathcal{F}_{\vec P}(n)\,,
\ee
as we have in \eqref{e210}. Let $
\log \mathcal{F}_{\vec P}(n) = -\frac c6 f_{\vec P}(n)\,.$ Then at the saddle point, using the large$-c$ approximation $\log S_{P_i 1} = 2\pi b^{-1} P_i$,\be\label{e235}
\log Z_n \sim 2\pi b^{-1} \sum_i P^*_{i,n} - b^{-2}f_{\vec P^*_n}(n) 
\ee
on the saddle-point values of the $\vec{P}$'s.
Plugging \eqref{e235} into \eqref{replica}, one can show that at large $c$,
\
\be\label{e236}
S \sim  2\pi b^{-1} \sum_i P^*_{i,n=1} + b^{-2}[\partial_n f_{\vec P_n^*} (n)|_{\vec P= \vec P^*_1, n=1} - f_{\vec{P}^*_{n=1}}(1)]\,.
\ee
Hence \eqref{sp1p} follows if the bracketed term in \eqref{e236} vanishes. A sufficient condition for this is  
\eqref{flin}.

\cite{Bao:2025plr} then asserts that \eqref{flin} is true for the same ladder-type  block as our $\mathcal{F}^{L(S/T)}_{P_1P_2}$. \footnote{See their eq. (4.40) and the discussion above their (4.33).}
An intuitive justification is that since  $\mathcal{F}(n)$ looks exactly like $n$ sewn-together copies of $\mathcal{F}(n=1)$, \eqref{flin} would follow as long as the blocks are well-approximated by a scalar action, which seems plausible at large $c$. 
Operationally, assuming replica symmetry, we can write each $\mathcal{F}_{\vec P}(n)$ as $\Tr(K_{\vec P}^n)$ for $K_{\vec P}$ the operator obtained by cutting the $Z_n$ path integral into $n$ identical building blocks. Then \eqref{flin} would follow as long as $K_{\vec P}$ was dominated by a leading eigenvalue at large $c$. 

We can't currently check this condition, again because we don't know $\mathcal{F}_{\vec P}(n)$
in closed form. However if we assume it, \eqref{sp1p} follows for the two-interval case to the same level of rigor as in \cite{Bao:2025plr}.

\subsubsection{Excited state case for light operators at the origin of the Euclidean disk}

For the case of a single interval in an excited state set up by inserting a generic operator $\OO$ at the origin of the Euclidean disk, we likewise cannot explicitly show \eqref{sp1p} because the analog of $p_{\vec P}$ in this case, \eqref{ppipb}, contains a Virasoro block which is not currently known in closed form. We can therefore make the same indirect argument as in section \ref{s252}, but not check a core assumption for that argument or
	directly evaluate \eqref{sp1p}.

More precisely, the analog of $p_{\vec P}$ in this case contains (after doubling)
a Virasoro block with two operator insertions of effective weight $h_\OO$ on a torus of length $2\ell$ and width $2\pi$, separated by a distance $L$ along the short cycle of the torus, and carrying Liouville momenta $P_i$, $P_b$ along the segments of length $L$ and $2\pi-L$ respectively. (See Figure \ref{fig2}.) For generic heavy $h_\OO$, this object is not known in closed form.

When $h_\OO$ is parametrically light, however -- more precisely, in the limit $h_\OO^2$, $\delta h^2 \ll h_i, h_b, c$ for $\delta h = h_b - h_i$ -- the descendants are suppressed and the necklace block is well-approximated by a product of Boltzmann factors \cite{Fitzpatrick:2014vua}, 
\be\label{smallh}
\mathcal{F}^{N(S)}_{P_i,P_b}(n=1) \sim e^{-L\frac{\pi}{\ell}P_i^2} e^{-(2\pi -L)\frac \pi \ell P_b^2}\,.
\ee
This gives a closed-form expression for $p_{PiP_b}$, \eqref{ppipb}, that we can use to find its peak. 

To find it, we  need to massage the $C_0(P_i,P_b,P_\OO)$, \eqref{c0}, inside the fusion kernel to a form that we can work with. 
This can be done by splitting $C_0$ into separate terms depending on how they scale with $b$, then using the identities 
\be
\Gamma_b(by) \sim \frac{\Gamma_b(Q)}{2\pi b}(2\pi b^3)^{y/2}\Gamma(y)\,, \qquad \Gamma_b(nQ + by) \sim \Gamma_b(nQ)\left(\frac{\sqrt{2\pi}b^{n-1/2}}{\Gamma(n)} \right)^y\,,
\ee
for $b \rightarrow 0$ with fixed $y$ and integer $n$ \cite{Ghosh:2019rcj}, and
\be
	\log \Gamma_b\left(b^{-1} x + \frac b 2 \right) \sim -\frac 1{b^2}\int_{\frac 12}^x dt \log\Gamma(t) + \frac 1{2b^2}\left( \frac 12 -x \right)^2\log b + \frac{2x-1}{4b^2}\log 2\pi + O(c^0)
\ee
for $x$ of order 1 \cite{Collier:2018exn}, to simplify them.
In the limit of light $h_\OO$, this yields
\be\label{pplight}
p_{P_iP_b} \sim S_{P_b1}\, S_{P_i1}\, e^{- L \frac{\pi}{\ell}P_i^2} e^{-(2\pi-L)\frac{\pi}{\ell}P_b^2}\,\frac{\Gamma(2h_\OO \pm ib^{-1}(P_i - P_b))}{\Gamma(4h_\OO)}(2h_\OO \pm ib^{-1}(P_i + P_b))^{2h_\OO\, \pm\, ib^{-1}(P_i + P_b)}
\ee
up to a $P_i$, $P_b$-independent prefactor. 

We'd like to extremize \eqref{pplight} w.r.t. $P_i$ and $P_b$. But this combination was extremized earlier in the literature, where it appeared as the large-$k$ limit of the integrand in the Euclidean Schwarzian two-point function \cite{Goel:2018ubv}. 
\footnote{
Namely, the Euclidean Schwarzian two-point function on a thermal circle of circumference $\beta$ reads
\be\label{schwarzian}
\langle \OO(\tau) \OO(0)\rangle_\beta = \prod_{i = 1,2} dk_i\, \rho(k_i) e^{-\frac{k_i^2}{2C}\tau} e^{-\frac{k_2^2}{2C}(\beta - \tau)}\frac{\Gamma(\ell \pm ik_1 \pm ik_2)}{\Gamma(2\ell)}
\ee
for $\rho(k_i) = k_i \sinh 2\pi k_i$. With the map $P_i = bk_1$, $P_b = bk_2$,  $\tau = L$, $\beta = 2\pi$, $C = \frac{\ell}{2\pi b^2}$ and $\ell = 2h_\OO$, this matches \eqref{pplight} at large $k$ (in particular, once we apply Stirling's formula to $\Gamma(\ell \pm i(k_1 + k_2)$) in \eqref{schwarzian}.) The result of extremizing \eqref{schwarzian} at light $h_\OO$ appears around eq. (3.28) of \cite{Goel:2018ubv}, where $\tau$ in this part of their paper was rescaled by an extra 
	1/2 (i.e. to compare their (3.28) with us one should use $\tau = L/2$). 
}
Evaluating $\log S_{P_b 1}$ on their result, we find 
\be\label{esaddle}
\left. \log S_{P_b 1}\right|_{P_b = P_b^*} = \frac{c\ell}{6} +  4h_\OO\left(1 - \frac L 2 \cot \frac L 2\right)\,.
\ee

On the other hand, the entanglement entropy of a single interval in an excited state set up by putting an operator $\OO$ of arbitrary weight at the origin of the Euclidean disk was computed using twist operators in \cite{Asplund:2014coa}. Those authors found that 
\be\label{asp}
S = \frac c3 \log\left(\frac{\beta_\OO}{\pi\epsilon} \sinh \left(\frac{L\pi}{\beta_\OO} \right) \right)\,,
\ee
for
\be
\beta_\OO = \frac{2\pi}{\sqrt{24h_\OO/c-1}}\,
\ee
(again assuming that $L < 2\pi -L$; otherwise we should swap $L <> 2\pi -L$).
In the light limit, expanding \eqref{asp} in $h_\OO/c$, 
\footnote{ See also eq. (1.3) of \cite{Belin:2021htw} who independently computed the light-$h_\OO$ limit of the entanglement entropy in this case, with  my $L$ = their $\theta$.
}
\be\label{smalloe}
S \sim  {\frac c 3 \log \left(\frac 2 \epsilon \sin \left(\frac L 2  \right) \right)} + 4h_\OO - 2Lh_\OO \cot \frac L 2 + O(h_\OO/c)\,.
\ee Comparing \eqref{esaddle} to \eqref{smalloe},  we find a match.

\section{Boundary interpretation of the cross-channel entropy formula}\label{s3}

Next, we interpret \eqref{rc} from a microscopic/boundary point of view.

\subsection{Eq. \eqref{rc} resembles an algebraic entanglement entropy with sector multiplicities
for an algebra $\AAA_{IR}$ with center labeled by $\vec{P}$} \label{s31}

Our first interpretative remark is that \eqref{rc} suggests the entropy may be organized by an underlying subalgebra $\AAA_{IR}$ of the bounded operator algebra acting on region $A$, with a center labeled by $\vec{P}$. In particular, the first (Shannon) and third (IR) terms in \eqref{rc} resemble an algebraic entanglement entropy for such an $\AAA_{IR}$.

Suppose there is such an $\AAA_{IR}$ with center labeled by $\vec{P}$. By definition \footnote{This brief exposition follows section III.B of the review article \cite{Liu:2025aa}.},  there then exists a decomposition of the Hilbert space on region $A$ 
\be
\HH_A = \bigoplus_{\vec P}( \HH^{IR}_{\vec P} \otimes \HH^{UV}_{\vec P})
\ee
for which 
\be
\AAA_{IR} = \bigoplus_{\vec P}(B(\HH^{IR}_{\vec P}) \otimes {\bf1}^{UV}_{\vec P})\,,
\ee
where $B(\HH)$ are the bounded operators acting on $\HH$. I.e., operators in $\AAA_{IR}$ cannot change the center label $\vec{P}$ and act on $\HH_{\vec P}^{IR}$ within a fixed $\vec{P}$-sector, while an optional per-sector multiplicity factor  $\HH^{UV}_{\vec P}$ is invisible to them. 

Let $\Pi_{\vec P}$ be the projection operator onto each $\vec{P}$-sector. 
For any state $\rho$ on $\HH_A$,  
let 
\begin{eqnarray}
p_{\vec P} &=& \Tr(\Pi_{\vec P}\,\rho), \label{defpp} \\
\rho_{\vec P} &=& p_{\vec P}^{-1}\Pi_{\vec P}\, \rho \, \Pi_{\vec P}\,, \label{defrhop} \\
\rho_{\vec P}^{IR} &=& \Tr_{\HH^{UV}_{\vec P}}(\rho_{\vec P})\,.
\end{eqnarray}
Then one can show that the restriction of $\rho$ to $\AAA_{IR}$ \footnote{I.e., the unique element $\rho_{\AAA_{IR}} \in \AAA_{IR}$ s.t.  $\tr_{\AAA_{IR}}(\rho_{\AAA_{IR}} A) = \Tr_{\HH}(\rho A)$ $\forall$ $A \in \AAA_{IR}$, where the trace on $\AAA_{IR}$ is the sum of the trace on each factor of the tensor decomposition,
	\be\label{tralg}
	\tr_{\AAA_{IR}}(A)  = \sum_{\vec P} \Tr_{\HH^{IR}_{\vec P}}(A_{\vec P}),  \qquad  A = \sum_{\vec P}(A_{\vec P} \otimes {\bf 1}^{UV}_{\vec P})\,.
	\ee
} 
is  
\be
\rho_{\AAA_{IR}} = \sum_{\vec P} p_{\vec P}(\rho^{IR}_{\vec P}\otimes {\bf 1}^{UV}_{\vec P})
\ee
and that the algebraic entanglement entropy of $\AAA_{IR}$ in state $\rho$ is 
\begin{eqnarray}
S_{\AAA_{IR}} &=& -\tr_{\AAA_{IR}} \rho_{\AAA_{IR}} \log \rho_{\AAA_{IR}}  \\ &=& -\sum_{\vec P} p_{\vec P} \log p_{\vec P} + \sum_{\vec P} p_{\vec P} S(\rho_{\vec P}^{IR}),  \label{algee}
\end{eqnarray}
where $\tr_{\AAA_{IR}}$ is defined as in \eqref{tralg}, and $ S(\rho_{\vec P}^{IR}) = -\Tr
	(\rho^{IR}_{\vec P} \log \rho^{IR}_{\vec P})$.
Hence, a Shannon term $-\sum_{\vec P} p_{\vec P} \log p_{\vec P}$  
and per-sector quantum entropy are generic consequences of a center labeled by $\vec{P}$.

\vspace{4mm}

In this framework, we can also account for the $\sum_i \log S_{P_i 1}$ term in \eqref{rc} by assuming that each $\vec{P}$-sector carries (a continuum analog of) a maximally mixed state of that dimension in the multiplicity space $\HH^{UV}_{\vec P}$. Suppose that $\rho$ is block-diagonal in $\vec{P}$ \footnote{
One generic way to enforce block-diagonality would be if $\rho$ itself was a reduced density matrix for an arbitrary state in the diagonal subspace $\HH = \bigoplus_{\vec P} \HH^A_{\vec P} \otimes \HH^{\bar A}_{\vec P}$ of a larger tensor product Hilbert space $\HH_A \otimes \HH_{\bar A}$, whose tensor factor $\HH_{\bar A}$ also had a matching decomposition $\HH_{\bar A} = \oplus_{\vec P} \HH^{\bar A}_{\vec P}$. In section \ref{s421} we'll argue that (when the global CFT is in the vacuum state) this is plausibly the physical mechanism giving rise to block-diagonal structure in a single-sided reduced density matrix $\rho$ on region $A$ underlying \eqref{rc}, with $\HH_{\bar{A}}$ the Hilbert space of the BCFT on region $\bar{A}$. 
} and factorizes within each block:
\be\label{e38}
		\rho = \bigoplus_{\vec P} p_{\vec P}\, \rho_{\vec P}, \qquad \rho_{\vec P} = \rho^{UV}_{\vec P} \otimes \rho^{IR}_{\vec P}\,.
\ee
Then the von Neumann entropy of $\rho$ would be 
\be\label{e39}
S(\rho) = -\sum_{\vec P} p_{\vec P} \log p_{\vec P} + \sum_{\vec P} p_{\vec P} \left(S(\rho_{\vec P}^{UV}) + S(\rho^{IR}_{\vec P})\right)\,.
\ee
If the UV factor is maximally mixed on a $\vec{P}$-sector-dependent but state-independent space of dimension $d_{\vec P}$, 
\be\label{uvf}
\rho^{UV}_{\vec P} = \frac{1}{d_{\vec P}} {\bf 1}^{UV}_{\vec P}, \qquad S(\rho^{UV}_{\vec P}) = \log d_{\vec P}\,,
\ee 
\eqref{e39} becomes
\be\label{e311}
S(\rho) = -\sum_{\vec P} p_{\vec P} \log p_{\vec P} + \sum_{\vec P} p_{\vec P}\left (\log d_{\vec P} + S(\rho^{IR}_{\vec P})\right) = S_{\AAA_{IR}} + \sum_{\vec P} p_{\vec P}\log d_{\vec P}\,,
\ee
matching \eqref{rc} if we identify  $\log d_{\vec P} = \sum_i \log S_{P_i 1}$. We will say more about how to make sense of the non-integer r.h.s. in section \ref{s33}.

\subsection{The Virasoro 
algebra in region $A$ 
is a natural candidate for $\AAA_{IR}$}\label{s32}

Conditioned on accepting the analogy of section \ref{s31}, the Virasoro/stress tensor algebra of the BCFT in region $A$ is a natural candidate for $\AAA_{IR}$.

The main piece of evidence is the construction of section \ref{s24}. There, for each case studied in section \ref{s2}, we constructed (up to a choice of Euclidean base point) a per-sector state whose von Neumann entropy reproduced $S_{IR}(\vec P)$ in \eqref{rc}. In the analogy of section \ref{s31}, this state should be identified with $\rho^{IR}_{\vec P}$ in \eqref{algee}, \eqref{e311}. It's therefore natural to identify the Hilbert space of the theory appearing in the construction of section \ref{s24} with the Hilbert space $\HH^{IR}_{\vec P}$ that supports the $\rho^{IR}_{\vec P}$'s in section \ref{s31}. 
But that theory was Virasoro TQFT, a theory that explicitly knows only about the Virasoro/stress tensor subalgebra of any would-be boundary CFT.

\subsection{The $\sum_i \log S_{P_i1}$ term can be explained by coarse-graining Virasoro primaries} \label{s33}

Next, {if} indeed  the Virasoro subalgebra in region $A$ is the organizing principle behind the Shannon/IR terms in \eqref{rc}, then the ``$\log S_{P_i 1}"$ term has a natural interpretation as the density-of-states part of a sector multiplicity arising after we coarse-grain the heavy primaries of the BCFT on region $A$  into bins labeled by Liouville momenta, with any coarse-graining scheme that meets the criteria outlined below \cite{Soni:2025qau}.

The motivation for this remark is that on the one hand, the BCFT on region $A$ in fact has the mathematical structure of a Virasoro subalgebra fibered over Virasoro primaries, but over a {discrete} collection of primaries, 
each with no innate large multiplicity.
On the other hand, the asymptotic density of states of heavy primaries in a holographic CFT is precisely 
 \be\label{hds}
 \rho (P) \sim S_{P1}\,.
 \ee 
 
For a concrete example of how coarse-graining the Shannon term of a fine-grained entropy could give rise to the $\log S_{P_i1}$ term in \eqref{rc}, suppose we start with a block-diagonal state 
\be\label{e313}
\rho = \bigoplus_{P_i \in P_d}  p_{P_i}\rho_{P_i}
\ee
with blocks labeled by discrete fine-grained labels $P_i \in P_d$. This state has the von Neumann entropy 
\be\label{e314}
S(\rho) = -\sum_{P_i \in P_d} p_{P_i} \log p_{P_i} + \sum_{P_i \in P_d} p_{P_i} S(\rho_{P_i})\,,
\ee
for $p_{P_i} = \Tr(\Pi_{P_i}\rho)$. Now partition the $P_i$'s into non-overlapping bins 
 \be
\mathcal{B}_P = \left[P - \frac{\delta P}{2}, P + \frac{\delta P}{2}
\right)\,,
\ee 
each centered at one of a discrete set of values $P$, for choices of the bin centers $P$ and widths $\delta P$ that are free up to satisfying criteria for a continuum limit that we'll state shortly below. Let the coarse-grained bin probability be  
\be
p_{\BB_P} = \sum_{P_i \in P_d \cap \mathcal{\BB}_P} p_{P_i} 
\ee
and let $n_{\BB_P}$ be the number of primaries in the bin $\BB_P$.
Moreover, suppose the weights within each bin are approximately flat:
\be \label{wflat}
p_{P_i} \sim \frac{p_{\BB_P}}{n_{\BB_P}} \qquad \forall\, P_i \in \mathcal{B}_P\,.
\ee
Then the discrete fine-grained Shannon entropy in \eqref{e314} reorganizes into a discrete coarse-grained Shannon entropy over the bins, plus a multiplicity term from forgetting which primary is being measured within each bin:
\be\label{s318}
 - \sum_{P_i \in P_d} p_{P_i} \log p_{P_i} \sim \sum_{\BB_P} -p_{\BB_P} \log p_{\BB_P} +  p_{\BB_P} \log n_{\BB_P}\,.
\ee

The latter will give a log(density of states) term in a continuum limit. To take such a limit, suppose each bin is large enough to contain many primaries, but small enough that would-be continuum densities $p_P$, $\rho(P)$ are approximately constant within each bin,   

 \be\label{33crit}
 \rho(P) \delta P \gg 1, \qquad \delta P ||\partial_P \log p_P || \ll 1\,, \qquad \delta P ||\partial_P \log \rho(P)||\ll1\,.
 \ee
 Then 
\begin{eqnarray}\label{continuum}
p_{\BB_P} &=& \int_{\BB_P} dP \sum_{P_i\in P_d} p_{P_i} \delta (P - P_i) \sim \int_{\BB_P} dP\, p_P \sim p_P\, \delta P\,, \label{e318} \\
n_{\BB_P} &=& \int_{\mathcal{B}_P} dP \sum_{P_i\in P_d} \delta(P - P_i) \sim  \int_{\mathcal{B}_P} dP \rho(P) \sim \rho(P) \delta P\,. \label{e319}
\end{eqnarray}
Substituting these into \eqref{s318} yields
\be\label{e321}
\sum_{\BB_P} -p_{\BB_P} \log p_{\BB_P} +  p_{\BB_P} \log n_{\BB_P}  \sim \int dP\, (-p_P \log p_P + p_P \log \rho(P))\,.
\ee
Eq. \eqref{e321} matches the first two terms of \eqref{rc}, given \eqref{hds} for a holographic CFT.

Comments:
\begin{itemize}
\item[(*)] In this example, the assumption \eqref{wflat} that the fine-grained sector weights are flat within each bin was made to keep the derivation as simple as possible, but is stronger than is actually needed for the argument to work. For the argument to work, we only need deviations from \eqref{wflat} to be small enough that \eqref{s318} is approximately true. 
\item[(*)] Here we discussed just the reorganization of the first term in \eqref{e314}. The second term has a coarse-grained reorganization as well, 	
		\be\label{s322} \sum_{P_i \in P_d} p_{P_i} S(\rho_{P_i}) \sim  \sum_{\BB_P}\,  p_{\BB_P} \left[\frac{1}{n_{\BB_P}}\sum_{P_i \in P_d \cap \BB_P} S(\rho_{P_i})\right]\,. \ee
	To fully recover \eqref{rc}, we would also need $S_{IR}(P)$ to match the continuum limit of the bracketed term on the r.h.s. of \eqref{s322}. This additional criterion could be satisfied with a similar smoothness condition as \eqref{33crit} for $S_{IR}(P)$ within each bin.
\item[(*)] 
The hard bin construction presented here is a simple example of a coarse-graining scheme.
More general coarse-graining schemes, e.g. with overlapping window functions, could also give \eqref{rc} in a continuum limit if they satisfy similar semiclassicality and smoothness conditions. Some such schemes are discussed in \cite{Soni:2025qau}.
\item[(*)] 
Returning to the present example, any choice of partition satisfying \eqref{s318}, \eqref{33crit} would generate a match between the fine-grained Shannon entropy of the microstates and a coarse-grained entropy with the functional form of \eqref{rc}. 
These criteria could be satisfied with qualitatively different scales for the bin width $\delta P$, with different physical implications. However, consistency of other aspects of the interpretation throughout Section \ref{s3} constrains the scale of coarse-graining allowed by a CFT that realizes the interpretation, beyond the requirements of \eqref{s318} and \eqref{33crit}.
\end{itemize}

\subsection{Positivity of bin probabilities constrains the live coarse-graining hypotheses}\label{s34}

Indeed, to close this section we note that \eqref{rc} contains within it quantitative criteria for the allowed scale of coarse-graining that would let the rest of the interpretation go through.

To recapitulate up to here: \eqref{rc} suggests, though does not prove, that entanglement entropy in holographic 2d CFT's is organized by Virasoro representation theory. The evidence for this claim is that it requires several a priori unrelated puzzle pieces to fit together: that \eqref{rc} has the functional form discussed in section \ref{s31} with the same $\vec{P}$-labels for each term; that $S_{IR}(\vec P)$ can be identified with the von Neumann entropy of a state in Virasoro TQFT; that $S_{P1}$ coincides with the density of states for heavy primaries in a holographic CFT; and that this same structure appears over the geometrically distinct situations studied in section \ref{s2}. (Moreover, other authors \cite{Casini:2019kex} previously conjectured that Virasoro representation theory may govern the organization of entanglement entropy in AdS$_3$/CFT$_2$, from somewhat different considerations. See footnote 18.) However, some of the arguments in sections \ref{s31} - \ref{s33} were at the level of analogies instead of explicit constructions.

An important fact that we have not yet used is that we know the actual density $p_{\vec P}$ appearing in \eqref{rc}, in \eqref{pp}, \eqref{pp1p2}, and \eqref{ppipb}. This provides a quantitative criterion that in principle could strengthen (or falsify) our interpretation: to show (or disprove) that for each actual holographic BCFT on the regulated region $A$, and reduced state $\rho$ on $A$, there exists a choice for how to coarse-grain the heavy primaries into bins $\BB$ s.t.
\footnote{
Throughout the main text of this section, our discussion assumes the hard bin coarse-graining from section \ref{s33}. Technically, fuzzier notions of coarse-graining as e.g. discussed in \cite{Soni:2025qau} could also be consistent with our interpretation, so the tests in \eqref{e324} - \eqref{e327} are logically representative but not the most stringent possible tests of our interpretation (i.e. they would be required for the hard bin coarse-graining presented in section \ref{s33} to work, but not for more general types of coarse-graining to work). In particular, self-consistency of our interpretation could also be achieved if \eqref{e324} were relaxed to positivity of more general finite-resolution smearings of $p_{\vec P}$, 
\be\label{fcg}
\int dP f_\alpha(P)\, p_P^{(1.2)} \geq 0\,,
\ee
for the overlapping windows $f_\alpha$ discussed in \cite{Soni:2025qau}. 
The point of this section is to emphasize that criteria like \eqref{fcg} - \eqref{e327} provides  consistency tests of our picture in principle, not to specify the exact strongest form of such tests.}
\footnote{For the excited state case in particular, another relaxation of \eqref{q1} that would preserve our broader interpretation would be to marginalize over $P_i$ and treat $P_b$ alone as a center variable, with $p_{P_b} = \int dP_i p_{P_iP_b}$ alone needing to satisfy a condition like \eqref{q1} or subsequent positivity conditions like in \eqref{fcg}, \eqref{e324} - \eqref{e327}. See \cite{Lin:2021tb} for a related discussion.} 
\be\label{q1}
\int_\BB  d\vec{P}\, p_{\vec P}^{(1.2)} \stackrel{?}{=} \Tr\, (\Pi_{\mathcal B} \rho)\,,
\ee
for $p_{\vec P}^{(1.2)}$ the explicit densities appearing in \eqref{pp}, \eqref{pp1p2}, and \eqref{ppipb}. 
Eq. \eqref{q1} is formal in that we don't currently have control over the microscopic spectrum of each BCFT with which to construct the right-hand projectors. However, weaker versions of \eqref{q1} where we relax the r.h.s. could potentially be used in the near term to constrain the space of coarse-grainings that would let the rest of the interpretation go through.

For example, the coarse-grained versions of $p_{\vec P}$ that constitute the would-be physical bin probabilities  $p_{\BB_{\vec P}}$ upon running \eqref{continuum} in reverse must be nonnegative, 
\be\label{e324}
\int_{\BB_{\vec P}} d\vec{P}\, p_{\vec P} \stackrel{?}{\geq} 0\,,
\ee
in order to admit any probabilistic interpretation en route to satisfying \eqref{q1}. This is noteworthy because the densities $p_{\vec P}$ in \eqref{pp1p2}, \eqref{ppipb} are not manifestly positive. In particular, they depend on conformal blocks that are not manifestly positive, with the explicit form of \eqref{e324} after plugging in \eqref{pp1p2} and \eqref{ppipb} respectively being  
\footnote{In these equations, we dropped a normalization constant $Z_1^{-1}$ because only the sign is being tested.}
\begin{eqnarray}
\int_{\BB_{\vec P}} dP_1 dP_2\, \,  S_{P_1 1}S_{P_2 1}\mathcal{F}^{L(S/T)}_{P_1P_2}(1) &\stackrel{?}{\geq}& 0\,, \label{e326} \\
\int_{\BB_{\vec P}} dP_i dP_b\, \,  S_{P_b 1}\, \mathbb{F}_{P_i 1}\left[\begin{tabular}{cc} $P_\OO$ & $P_b$ \\ $P_\OO$ & $P_b$ \end{tabular} \right] \mathcal{F}^{N(S/T)}_{P_i, P_b}(1)&\stackrel{?}{\geq}& 0\,. \label{e327}
\end{eqnarray}
The positivity conditions \eqref{e326}, \eqref{e327} (or more general analogs described in footnotes 14 and 15) could in principle be true (i) pointwise, (ii) only after averaging over $\delta P$-sized bins for a derivable-in-principle scale of $\delta P$ that satisfies the other requirements of section \ref{s33}, or (iii) not true for any $\delta P$ that satisfies the requirements of section \ref{s33}. Option (iii) would falsify our interpretation (modulo the caveats in footnotes 14 and 15).
On the other hand, if one independently assigns credence to our interpretation in this section and the next (based e.g. on the coincidences described at the top of this section), then 
	perhaps further analysis of such positivity criteria could help shed light on the type of coarse-graining required for the emergence of a semiclassical bulk in holography.

\section{Implications for holography}\label{s4}

So far, everything we discussed was for the boundary CFT. In this section, we discuss implications for holography.

\subsection{The RT area is the $\sum_i \log S_{P_i 1}$ term in \eqref{rc}}

Actually, there is one main result which is the punchline of this work.
In holographic CFT's, the entanglement entropy of a boundary region $A$ is represented in the bulk by the Ryu-Takayanagi formula, \eqref{rt}. A longstanding open question is to understand how the boundary physics organizes geometrically to yield the RT formula at large $c$, in particular what the area term is counting. 
	The main lesson of this work is that \eqref{rc} provides a quantitative bridge between the RT formula and a microscopic explanation for it implied by the discussion of section \ref{s3}. The bridge is that when we compare \eqref{rt} and \eqref{rc} term-by-term in $c$, the $O(c)$ area term can be associated with the $\log S_{P1}$ part of the dominant $\vec{P}$-sector in the boundary CFT, 
\footnote{The other, $O(c^0)$ terms contribute to but do not fully explain the FLM contribution, since we dropped non-universal contributions from light bulk matter when we made the vacuum block dominance approximation.}
\be\label{rtm} \left.\frac{\mbox{Area}(\gamma_A)}{4G_N} = \sum_i \log S_{P_i1}\right|_{\vec{P}=\vec{P}^*}\,.
\ee
Hence, the area term counts ``whatever the $\log S_{P_i 1}$ term is counting".

\vspace{4mm} 
In Section \ref{s3}, we then argued that from a boundary point of view, the $\log S_{P_i 1}$ term has the interpretation of arising from the density-of-states part of a count of heavy Virasoro primaries in the BCFT on region $A$, in a $\delta P$-sized window around a dominant Liouville momentum $P^*$ that depends on the global state and subregion of the entanglement calculation, for a coarse-graining scale $\delta P$ that satisfies the 
criteria of sections \ref{s33} and \ref{s34}. Given \eqref{rtm}, this is our candidate answer for what the RT area counts from a boundary point of view.

\subsection{Identification of prospective ``hardware" carrying the RT entropy}\label{s421}

This result points to a natural set of candidate physical carriers of the RT entropy.

From a two-sided point of view (i.e. on $A \cup \bar {A}$ instead of just region $A$), the regulated entanglement calculation in the CFT computes entanglement entropy on a state in the tensor product Hilbert space 
\be
\HH_A^{BCFT} \otimes \HH_{\bar A}^{BCFT} = \bigoplus_\alpha \mathcal{V}^A_\alpha \otimes \bigoplus_\beta \mathcal{V}^{\bar A}_\beta\,,
\ee
where $\VV_\alpha^A$, $\VV_{\beta}^{\bar A}$ are Virasoro modules in the respective BCFT's. However, states in the global CFT do not populate arbitrary pairs $\VV^A_\alpha \otimes \VV^{\bar A}_\beta$. Rather, a fixed global state in a representation $R$ only supports pairs whose fusion can produce that representation. Bilocal composite operators built from Virasoro-charged operators in $A$ and $\bar A$ whose representation labels are compatible with the global state, projected onto the representation of the global state,  are called Virasoro intertwiners. 

Given a two-sided description of the global state, the reduced density matrix on one side contains support on the Virasoro representation labels compatible with this global constraint. These labels are the would-be labels that appear in \eqref{rc} after coarse-graining, with the RT area being (the density-of-states part of) the log of the number of microscopic realizations of a $\delta P$-sized neighborhood of the dominant label $\vec{P}^*$, as described above.
\footnote{The idea that Virasoro representation theory could govern the organization of entanglement in holographic 2d CFT's was proposed previously in \cite{Casini:2019kex}, specifically using Virasoro intertwiners on $A \cup \bar A$. Based on an analogy between holographic mutual information and mutual information in a generic theory with superselection sectors, along with the observation that the Virasoro algebra  endows generic CFT$_2$'s with a superselection sector structure, those authors proposed that the RT area might equal a relative entropy between the global state on the regulated region  $A \cup \bar A$ and the same state after erasing the correlations of intertwiners on $A \cup \bar A$. Our result refines this conjecture by identifying the entropy with the log of the Cardy density of Virasoro primaries at a dominant Virasoro label.}

\vspace{4mm}
\noindent Eq. \eqref{rtm} also points to a closely related interpretation of what the RT area could mean from a ``pre-geometric"
 bulk point of view. Namely:

In earlier works (e.g. \cite{Harlow:2015lma}, \cite{Donnelly:2016auv},  \cite{Lin:2017uzr}) it was suggested, again qualitatively, that perhaps the RT area could be the gravitational analog of an ``edge term" that appears when one computes entanglement entropy in ordinary gauge theories in a way that keeps track of internal correlations of line operators that pass through the cut. (See e.g. \cite{Donnelly:2014gva}, \cite{Donnelly:2011hn}, \cite{Ghosh:2015iwa}.) 
In particular, \cite{Lin:2017uzr} observed that there is a formal equivalence between the area term in the RT formula and a ``$\log \dim R"$ contribution to entanglement entropy in a compact lattice gauge theory, whose role is to keep track of link holonomies labeled by the representation $R$ that cross the entangling boundary. \footnote{ Readers familiar with this topic may recognize that \eqref{rc} superficially resembles said formula for entanglement entropy in a compact lattice gauge theory. 
However, \eqref{rc} has the wrong counting for the analog of the $``\log \dim R"$ term to be interpreted as describing a $1+1d$ gauge theory on the boundary. In a $1+1d$ gauge theory,  one would assign a ``$\log \dim R$" term to each entangling boundary/endpoint of an interval \cite{Donnelly:2014gva}, while in \eqref{rc} we assign one ``$\log S_{P1}$" term to each disconnected region/{\it pair} of entangling boundaries. 
	This however has the right counting for the $\log S_{P1}$ terms to be edge-localized entropy at the ends of a line operator {\it in an emergent bulk}, as proposed in the main text.
}
Two observations let us upgrade this to a more quantitative proposal:

\begin{enumerate}
\item For a Chern-Simons gauge theory in particular, the natural edge contribution of a line operator in representation $R$ is the log of the Plancherel measure of the quantum group whose fusion/braiding data appears when we manipulate the Wilson lines of the theory. (See \cite{Mertens:2025ydx} for a recent discussion of this point.)
\item The modular S-matrix element appearing on the r.h.s. of \eqref{rc} coincides with the (modular double) Plancherel measure of the quantum group $U_q(sl2)$ \cite{Ponsot:1999uf},  \cite{Blommaert:2018iqz}. 
\end{enumerate}
The appearance of the same quantum group Plancherel measure in these separate contexts suggests that we view the RT area as an edge term for a noncompact Chern-Simons-like theory with the hidden quantum group $U_q(sl2)$ from a bulk point of view. In particular, it suggests that we identify a given RT area term with the edge term of a bulk line operator at the matching representation $P^*$ in \eqref{rtm}, which we interpret as the bulk dual of the coarse-grained boundary intertwiners in a neighborhood of $P^*$. This bulk proposal was basically already put forth in \cite{Mertens:2022ujr, Wong:2022eiu} for two-sided BTZ entropy and the RT formula for a single interval in the vacuum state in $AdS_3$. Our results suggest it may generalize to less symmetric situations.

\section{Discussion} \label{s5}

The story presented in this work relied on a few assumptions that we made along the way. One gap is to explicitly find the peak of $p_{\vec P}$ and use it to show that the $\sum_i \log S_{P_i 1}$ term dominates in \eqref{rc} for the multiple interval and heavy excited state cases, 
	instead of leaning on a saddle point assumption. However, as discussed in section \ref{s25}, this may be hard to do with current technology.

Another important gap is to  better understand whether the $p_{\vec P}$'s derived from our replica manipulations can be interpreted as actual Born-rule probabilities, perhaps after coarse-graining, as discussed in section \ref{s34}. 
As a stepping stone towards this goal, understanding when (a coarse-grained version of the) $p_{\vec P}$'s are positive, as in \eqref{fcg}, \eqref{e324} - \eqref{e327}, could perhaps constrain the amount of coarse-graining needed for our interpretation of the RT formula to work, and is probably the most concrete future direction suggested by this work.

It could be interesting to understand how our results change as we back away from the semiclassical limit. In general, $1/c$ or stringy corrections would require us to back away from the vacuum block dominance assumption of \cite{Hartman:2013mia}, which may be hard to do head-on. 
Perhaps a more tractable question is whether there are other special limits or situations in AdS$_3$/CFT$_2$ where the Euclidean partition function can be written in a way that would lead to an entropy formula like \eqref{rc}.

Finally, it would be interesting to understand how to generalize the structure in this paper to higher dimensions or to non-AdS backgrounds. However, the technical results in this paper depended heavily on crossing symmetry in the boundary 2d CFT, so such developments may have to await new technology.

\bibliographystyle{ssg}
\bibliography{bhe}		
\end{document}